\hsize=31pc
\vsize=49pc
\lineskip=0pt
\parskip=0pt plus 1pt
\hfuzz=1pt  
\vfuzz=2pt
\pretolerance=2500
\tolerance=5000
\vbadness=5000
\hbadness=5000
\widowpenalty=500
\clubpenalty=200
\brokenpenalty=500
\predisplaypenalty=200
\voffset=-1pc
\nopagenumbers     
\catcode`@=11
\newif\ifams
\amsfalse
%
%
%
%
\newfam\bdifam
\newfam\bsyfam
\newfam\bssfam
\newfam\msafam
\newfam\msbfam
\newif\ifxxpt   
\newif\ifxviipt 
\newif\ifxivpt  
\newif\ifxiipt  
\newif\ifxipt   
\newif\ifxpt    
\newif\ifixpt   
\newif\ifviiipt 
\newif\ifviipt  
\newif\ifvipt   
\newif\ifvpt    
%
%
\def\headsize#1#2{\def\headb@seline{#2}%
                \ifnum#1=20\def\HEAD{twenty}%
                           \def\smHEAD{twelve}%
                           \def\vsHEAD{nine}%
                           \ifxxpt\else\xdef\f@ntsize{\HEAD}%
                           \def\m@g{4}\def\s@ze{20.74}%
                           \loadheadfonts\xxpttrue\fi
                           \ifxiipt\else\xdef\f@ntsize{\smHEAD}%
                           \def\m@g{1}\def\s@ze{12}%
                           \loadxiiptfonts\xiipttrue\fi
                           \ifixpt\else\xdef\f@ntsize{\vsHEAD}%
                           \def\s@ze{9}%
                           \loadsmallfonts\ixpttrue\fi
                      \else
                \ifnum#1=17\def\HEAD{seventeen}%
                           \def\smHEAD{eleven}%
                           \def\vsHEAD{eight}%
                           \ifxviipt\else\xdef\f@ntsize{\HEAD}%
                           \def\m@g{3}\def\s@ze{17.28}%
                           \loadheadfonts\xviipttrue\fi
                           \ifxipt\else\xdef\f@ntsize{\smHEAD}%
                           \loadxiptfonts\xipttrue\fi
                           \ifviiipt\else\xdef\f@ntsize{\vsHEAD}%
                           \def\s@ze{8}%
                           \loadsmallfonts\viiipttrue\fi
                      \else\def\HEAD{fourteen}%
                           \def\smHEAD{ten}%
                           \def\vsHEAD{seven}%
                           \ifxivpt\else\xdef\f@ntsize{\HEAD}%
                           \def\m@g{2}\def\s@ze{14.4}%
                           \loadheadfonts\xivpttrue\fi
                           \ifxpt\else\xdef\f@ntsize{\smHEAD}%
                           \def\s@ze{10}%
                           \loadxptfonts\xpttrue\fi
                           \ifviipt\else\xdef\f@ntsize{\vsHEAD}%
                           \def\s@ze{7}%
                           \loadviiptfonts\viipttrue\fi
                \ifnum#1=14\else
                \message{Header size should be 20, 17 or 14 point
                              will now default to 14pt}\fi
                \fi\fi\headfonts}
%
%
\def\textsize#1#2{\def\textb@seline{#2}%
                 \ifnum#1=12\def\TEXT{twelve}%
                           \def\smTEXT{eight}%
                           \def\vsTEXT{six}%
                           \ifxiipt\else\xdef\f@ntsize{\TEXT}%
                           \def\m@g{1}\def\s@ze{12}%
                           \loadxiiptfonts\xiipttrue\fi
                           \ifviiipt\else\xdef\f@ntsize{\smTEXT}%
                           \def\s@ze{8}%
                           \loadsmallfonts\viiipttrue\fi
                           \ifvipt\else\xdef\f@ntsize{\vsTEXT}%
                           \def\s@ze{6}%
                           \loadviptfonts\vipttrue\fi
                      \else
                \ifnum#1=11\def\TEXT{eleven}%
                           \def\smTEXT{seven}%
                           \def\vsTEXT{five}%
                           \ifxipt\else\xdef\f@ntsize{\TEXT}%
                           \def\s@ze{11}%
                           \loadxiptfonts\xipttrue\fi
                           \ifviipt\else\xdef\f@ntsize{\smTEXT}%
                           \loadviiptfonts\viipttrue\fi
                           \ifvpt\else\xdef\f@ntsize{\vsTEXT}%
                           \def\s@ze{5}%
                           \loadvptfonts\vpttrue\fi
                      \else\def\TEXT{ten}%
                           \def\smTEXT{seven}%
                           \def\vsTEXT{five}%
                           \ifxpt\else\xdef\f@ntsize{\TEXT}%
                           \loadxptfonts\xpttrue\fi
                           \ifviipt\else\xdef\f@ntsize{\smTEXT}%
                           \def\s@ze{7}%
                           \loadviiptfonts\viipttrue\fi
                           \ifvpt\else\xdef\f@ntsize{\vsTEXT}%
                           \def\s@ze{5}%
                           \loadvptfonts\vpttrue\fi
                \ifnum#1=10\else
                \message{Text size should be 12, 11 or 10 point
                              will now default to 10pt}\fi
                \fi\fi\textfonts}
%
%
\def\smallsize#1#2{\def\smallb@seline{#2}%
                 \ifnum#1=10\def\SMALL{ten}%
                           \def\smSMALL{seven}%
                           \def\vsSMALL{five}%
                           \ifxpt\else\xdef\f@ntsize{\SMALL}%
                           \loadxptfonts\xpttrue\fi
                           \ifviipt\else\xdef\f@ntsize{\smSMALL}%
                           \def\s@ze{7}%
                           \loadviiptfonts\viipttrue\fi
                           \ifvpt\else\xdef\f@ntsize{\vsSMALL}%
                           \def\s@ze{5}%
                           \loadvptfonts\vpttrue\fi
                       \else
                 \ifnum#1=9\def\SMALL{nine}%
                           \def\smSMALL{six}%
                           \def\vsSMALL{five}%
                           \ifixpt\else\xdef\f@ntsize{\SMALL}%
                           \def\s@ze{9}%
                           \loadsmallfonts\ixpttrue\fi
                           \ifvipt\else\xdef\f@ntsize{\smSMALL}%
                           \def\s@ze{6}%
                           \loadviptfonts\vipttrue\fi
                           \ifvpt\else\xdef\f@ntsize{\vsSMALL}%
                           \def\s@ze{5}%
                           \loadvptfonts\vpttrue\fi
                       \else
                           \def\SMALL{eight}%
                           \def\smSMALL{six}%
                           \def\vsSMALL{five}%
                           \ifviiipt\else\xdef\f@ntsize{\SMALL}%
                           \def\s@ze{8}%
                           \loadsmallfonts\viiipttrue\fi
                           \ifvipt\else\xdef\f@ntsize{\smSMALL}%
                           \def\s@ze{6}%
                           \loadviptfonts\vipttrue\fi
                           \ifvpt\else\xdef\f@ntsize{\vsSMALL}%
                           \def\s@ze{5}%
                           \loadvptfonts\vpttrue\fi
                 \ifnum#1=8\else\message{Small size should be 10, 9 or 
                            8 point will now default to 8pt}\fi
                \fi\fi\smallfonts}
\def\F@nt{\expandafter\font\csname}
\def\Sk@w{\expandafter\skewchar\csname}
\def\@nd{\endcsname}
\def\@step#1{ scaled \magstep#1}
\def\@half{ scaled \magstephalf}
\def\@t#1{ at #1pt}
%
%
\def\loadheadfonts{\bigf@nts
\F@nt \f@ntsize bdi\@nd=cmmib10 \@t{\s@ze}%
\Sk@w \f@ntsize bdi\@nd='177
\F@nt \f@ntsize bsy\@nd=cmbsy10 \@t{\s@ze}%
\Sk@w \f@ntsize bsy\@nd='60
\F@nt \f@ntsize bss\@nd=cmssbx10 \@t{\s@ze}}
%
%
\def\loadxiiptfonts{\bigf@nts
\F@nt \f@ntsize bdi\@nd=cmmib10 \@step{\m@g}%
\Sk@w \f@ntsize bdi\@nd='177
\F@nt \f@ntsize bsy\@nd=cmbsy10 \@step{\m@g}%
\Sk@w \f@ntsize bsy\@nd='60
\F@nt \f@ntsize bss\@nd=cmssbx10 \@step{\m@g}}
%
%
\def\loadxiptfonts{%
\font\elevenrm=cmr10 \@half
\font\eleveni=cmmi10 \@half
\skewchar\eleveni='177
\font\elevensy=cmsy10 \@half
\skewchar\elevensy='60
\font\elevenex=cmex10 \@half
\font\elevenit=cmti10 \@half
\font\elevensl=cmsl10 \@half
\font\elevenbf=cmbx10 \@half
\font\eleventt=cmtt10 \@half
\ifams\font\elevenmsa=msam10 \@half
\font\elevenmsb=msbm10 \@half\else\fi
\font\elevenbdi=cmmib10 \@half
\skewchar\elevenbdi='177
\font\elevenbsy=cmbsy10 \@half
\skewchar\elevenbsy='60
\font\elevenbss=cmssbx10 \@half}
%
%
\def\loadxptfonts{%
\font\tenbdi=cmmib10
\skewchar\tenbdi='177
\font\tenbsy=cmbsy10 
\skewchar\tenbsy='60
\ifams\font\tenmsa=msam10 
\font\tenmsb=msbm10\else\fi
\font\tenbss=cmssbx10}%
%
%
\def\loadsmallfonts{\smallf@nts
\ifams
\F@nt \f@ntsize ex\@nd=cmex\s@ze
\else
\F@nt \f@ntsize ex\@nd=cmex10\fi
\F@nt \f@ntsize it\@nd=cmti\s@ze
\F@nt \f@ntsize sl\@nd=cmsl\s@ze
\F@nt \f@ntsize tt\@nd=cmtt\s@ze}
%
%
\def\loadviiptfonts{%
\font\sevenit=cmti7
\font\sevensl=cmsl8 at 7pt
\ifams\font\sevenmsa=msam7 
\font\sevenmsb=msbm7
\font\sevenex=cmex7
\font\sevenbsy=cmbsy7
\font\sevenbdi=cmmib7\else
\font\sevenex=cmex10
\font\sevenbsy=cmbsy10 at 7pt
\font\sevenbdi=cmmib10 at 7pt\fi
\skewchar\sevenbsy='60
\skewchar\sevenbdi='177
\font\sevenbss=cmssbx10 at 7pt}%
%
%
\def\loadviptfonts{\smallf@nts
\ifams\font\sixex=cmex7 at 6pt\else
\font\sixex=cmex10\fi
\font\sixit=cmti7 at 6pt}
%
%
\def\loadvptfonts{%
\font\fiveit=cmti7 at 5pt
\ifams\font\fiveex=cmex7 at 5pt
\font\fivebdi=cmmib5
\font\fivebsy=cmbsy5
\font\fivemsa=msam5 
\font\fivemsb=msbm5\else
\font\fiveex=cmex10
\font\fivebdi=cmmib10 at 5pt
\font\fivebsy=cmbsy10 at 5pt\fi
\skewchar\fivebdi='177
\skewchar\fivebsy='60
\font\fivebss=cmssbx10 at 5pt}
\def\bigf@nts{%
\F@nt \f@ntsize rm\@nd=cmr10 \@step{\m@g}%
\F@nt \f@ntsize i\@nd=cmmi10 \@step{\m@g}%
\Sk@w \f@ntsize i\@nd='177
\F@nt \f@ntsize sy\@nd=cmsy10 \@step{\m@g}%
\Sk@w \f@ntsize sy\@nd='60
\F@nt \f@ntsize ex\@nd=cmex10 \@step{\m@g}%
\F@nt \f@ntsize it\@nd=cmti10 \@step{\m@g}%
\F@nt \f@ntsize sl\@nd=cmsl10 \@step{\m@g}%
\F@nt \f@ntsize bf\@nd=cmbx10 \@step{\m@g}%
\F@nt \f@ntsize tt\@nd=cmtt10 \@step{\m@g}%
\ifams
\F@nt \f@ntsize msa\@nd=msam10 \@step{\m@g}%
\F@nt \f@ntsize msb\@nd=msbm10 \@step{\m@g}\else\fi}
\def\smallf@nts{%
\F@nt \f@ntsize rm\@nd=cmr\s@ze
\F@nt \f@ntsize i\@nd=cmmi\s@ze 
\Sk@w \f@ntsize i\@nd='177
\F@nt \f@ntsize sy\@nd=cmsy\s@ze
\Sk@w \f@ntsize sy\@nd='60
\F@nt \f@ntsize bf\@nd=cmbx\s@ze 
\ifams
\F@nt \f@ntsize bdi\@nd=cmmib\s@ze 
\F@nt \f@ntsize bsy\@nd=cmbsy\s@ze 
\F@nt \f@ntsize msa\@nd=msam\s@ze 
\F@nt \f@ntsize msb\@nd=msbm\s@ze
\else
\F@nt \f@ntsize bdi\@nd=cmmib10 \@t{\s@ze}%
\F@nt \f@ntsize bsy\@nd=cmbsy10 \@t{\s@ze}\fi 
\Sk@w \f@ntsize bdi\@nd='177
\Sk@w \f@ntsize bsy\@nd='60
\F@nt \f@ntsize bss\@nd=cmssbx10 \@t{\s@ze}}%
%
%
\def\headfonts{%
\textfont0=\csname\HEAD rm\@nd        
\scriptfont0=\csname\smHEAD rm\@nd
\scriptscriptfont0=\csname\vsHEAD rm\@nd
\def\rm{\fam0\csname\HEAD rm\@nd
\def\sc{\csname\smHEAD rm\@nd}}%
\textfont1=\csname\HEAD i\@nd         
\scriptfont1=\csname\smHEAD i\@nd
\scriptscriptfont1=\csname\vsHEAD i\@nd
\textfont2=\csname\HEAD sy\@nd        
\scriptfont2=\csname\smHEAD sy\@nd
\scriptscriptfont2=\csname\vsHEAD sy\@nd
\textfont3=\csname\HEAD ex\@nd        
\scriptfont3=\csname\smHEAD ex\@nd
\scriptscriptfont3=\csname\smHEAD ex\@nd
\textfont\itfam=\csname\HEAD it\@nd   
\scriptfont\itfam=\csname\smHEAD it\@nd
\scriptscriptfont\itfam=\csname\vsHEAD it\@nd
\def\it{\fam\itfam\csname\HEAD it\@nd
\def\sc{\csname\smHEAD it\@nd}}%
\textfont\slfam=\csname\HEAD sl\@nd   
\def\sl{\fam\slfam\csname\HEAD sl\@nd
\def\sc{\csname\smHEAD sl\@nd}}%
\textfont\bffam=\csname\HEAD bf\@nd   
\scriptfont\bffam=\csname\smHEAD bf\@nd
\scriptscriptfont\bffam=\csname\vsHEAD bf\@nd
\def\bf{\fam\bffam\csname\HEAD bf\@nd
\def\sc{\csname\smHEAD bf\@nd}}%
\textfont\ttfam=\csname\HEAD tt\@nd   
\def\tt{\fam\ttfam\csname\HEAD tt\@nd}%
\textfont\bdifam=\csname\HEAD bdi\@nd 
\scriptfont\bdifam=\csname\smHEAD bdi\@nd
\scriptscriptfont\bdifam=\csname\vsHEAD bdi\@nd
\def\bdi{\fam\bdifam\csname\HEAD bdi\@nd}%
\textfont\bsyfam=\csname\HEAD bsy\@nd 
\scriptfont\bsyfam=\csname\smHEAD bsy\@nd
\def\bsy{\fam\bsyfam\csname\HEAD bsy\@nd}%
\textfont\bssfam=\csname\HEAD bss\@nd 
\scriptfont\bssfam=\csname\smHEAD bss\@nd
\scriptscriptfont\bssfam=\csname\vsHEAD bss\@nd
\def\bss{\fam\bssfam\csname\HEAD bss\@nd}%
\ifams
\textfont\msafam=\csname\HEAD msa\@nd 
\scriptfont\msafam=\csname\smHEAD msa\@nd
\scriptscriptfont\msafam=\csname\vsHEAD msa\@nd
\textfont\msbfam=\csname\HEAD msb\@nd 
\scriptfont\msbfam=\csname\smHEAD msb\@nd
\scriptscriptfont\msbfam=\csname\vsHEAD msb\@nd
\else\fi
\normalbaselineskip=\headb@seline pt%
\setbox\strutbox=\hbox{\vrule height.7\normalbaselineskip 
depth.3\baselineskip width0pt}%
\def\sc{\csname\smHEAD rm\@nd}\normalbaselines\bf}
%
%
\def\textfonts{%
\textfont0=\csname\TEXT rm\@nd        
\scriptfont0=\csname\smTEXT rm\@nd
\scriptscriptfont0=\csname\vsTEXT rm\@nd
\def\rm{\fam0\csname\TEXT rm\@nd
\def\sc{\csname\smTEXT rm\@nd}}%
\textfont1=\csname\TEXT i\@nd         
\scriptfont1=\csname\smTEXT i\@nd
\scriptscriptfont1=\csname\vsTEXT i\@nd
\textfont2=\csname\TEXT sy\@nd        
\scriptfont2=\csname\smTEXT sy\@nd
\scriptscriptfont2=\csname\vsTEXT sy\@nd
\textfont3=\csname\TEXT ex\@nd        
\scriptfont3=\csname\smTEXT ex\@nd
\scriptscriptfont3=\csname\smTEXT ex\@nd
\textfont\itfam=\csname\TEXT it\@nd   
\scriptfont\itfam=\csname\smTEXT it\@nd
\scriptscriptfont\itfam=\csname\vsTEXT it\@nd
\def\it{\fam\itfam\csname\TEXT it\@nd
\def\sc{\csname\smTEXT it\@nd}}%
\textfont\slfam=\csname\TEXT sl\@nd   
\def\sl{\fam\slfam\csname\TEXT sl\@nd
\def\sc{\csname\smTEXT sl\@nd}}%
\textfont\bffam=\csname\TEXT bf\@nd   
\scriptfont\bffam=\csname\smTEXT bf\@nd
\scriptscriptfont\bffam=\csname\vsTEXT bf\@nd
\def\bf{\fam\bffam\csname\TEXT bf\@nd
\def\sc{\csname\smTEXT bf\@nd}}%
\textfont\ttfam=\csname\TEXT tt\@nd   
\def\tt{\fam\ttfam\csname\TEXT tt\@nd}%
\textfont\bdifam=\csname\TEXT bdi\@nd 
\scriptfont\bdifam=\csname\smTEXT bdi\@nd
\scriptscriptfont\bdifam=\csname\vsTEXT bdi\@nd
\def\bdi{\fam\bdifam\csname\TEXT bdi\@nd}%
\textfont\bsyfam=\csname\TEXT bsy\@nd 
\scriptfont\bsyfam=\csname\smTEXT bsy\@nd
\def\bsy{\fam\bsyfam\csname\TEXT bsy\@nd}%
\textfont\bssfam=\csname\TEXT bss\@nd 
\scriptfont\bssfam=\csname\smTEXT bss\@nd
\scriptscriptfont\bssfam=\csname\vsTEXT bss\@nd
\def\bss{\fam\bssfam\csname\TEXT bss\@nd}%
\ifams
\textfont\msafam=\csname\TEXT msa\@nd 
\scriptfont\msafam=\csname\smTEXT msa\@nd
\scriptscriptfont\msafam=\csname\vsTEXT msa\@nd
\textfont\msbfam=\csname\TEXT msb\@nd 
\scriptfont\msbfam=\csname\smTEXT msb\@nd
\scriptscriptfont\msbfam=\csname\vsTEXT msb\@nd
\else\fi
\normalbaselineskip=\textb@seline pt
\setbox\strutbox=\hbox{\vrule height.7\normalbaselineskip 
depth.3\baselineskip width0pt}%
\everymath{}%
\def\sc{\csname\smTEXT rm\@nd}\normalbaselines\rm}
%
%
\def\smallfonts{%
\textfont0=\csname\SMALL rm\@nd        
\scriptfont0=\csname\smSMALL rm\@nd
\scriptscriptfont0=\csname\vsSMALL rm\@nd
\def\rm{\fam0\csname\SMALL rm\@nd
\def\sc{\csname\smSMALL rm\@nd}}%
\textfont1=\csname\SMALL i\@nd         
\scriptfont1=\csname\smSMALL i\@nd
\scriptscriptfont1=\csname\vsSMALL i\@nd
\textfont2=\csname\SMALL sy\@nd        
\scriptfont2=\csname\smSMALL sy\@nd
\scriptscriptfont2=\csname\vsSMALL sy\@nd
\textfont3=\csname\SMALL ex\@nd        
\scriptfont3=\csname\smSMALL ex\@nd
\scriptscriptfont3=\csname\smSMALL ex\@nd
\textfont\itfam=\csname\SMALL it\@nd   
\scriptfont\itfam=\csname\smSMALL it\@nd
\scriptscriptfont\itfam=\csname\vsSMALL it\@nd
\def\it{\fam\itfam\csname\SMALL it\@nd
\def\sc{\csname\smSMALL it\@nd}}%
\textfont\slfam=\csname\SMALL sl\@nd   
\def\sl{\fam\slfam\csname\SMALL sl\@nd
\def\sc{\csname\smSMALL sl\@nd}}%
\textfont\bffam=\csname\SMALL bf\@nd   
\scriptfont\bffam=\csname\smSMALL bf\@nd
\scriptscriptfont\bffam=\csname\vsSMALL bf\@nd
\def\bf{\fam\bffam\csname\SMALL bf\@nd
\def\sc{\csname\smSMALL bf\@nd}}%
\textfont\ttfam=\csname\SMALL tt\@nd   
\def\tt{\fam\ttfam\csname\SMALL tt\@nd}%
\textfont\bdifam=\csname\SMALL bdi\@nd 
\scriptfont\bdifam=\csname\smSMALL bdi\@nd
\scriptscriptfont\bdifam=\csname\vsSMALL bdi\@nd
\def\bdi{\fam\bdifam\csname\SMALL bdi\@nd}%
\textfont\bsyfam=\csname\SMALL bsy\@nd 
\scriptfont\bsyfam=\csname\smSMALL bsy\@nd
\def\bsy{\fam\bsyfam\csname\SMALL bsy\@nd}%
\textfont\bssfam=\csname\SMALL bss\@nd 
\scriptfont\bssfam=\csname\smSMALL bss\@nd
\scriptscriptfont\bssfam=\csname\vsSMALL bss\@nd
\def\bss{\fam\bssfam\csname\SMALL bss\@nd}%
\ifams
\textfont\msafam=\csname\SMALL msa\@nd 
\scriptfont\msafam=\csname\smSMALL msa\@nd
\scriptscriptfont\msafam=\csname\vsSMALL msa\@nd
\textfont\msbfam=\csname\SMALL msb\@nd 
\scriptfont\msbfam=\csname\smSMALL msb\@nd
\scriptscriptfont\msbfam=\csname\vsSMALL msb\@nd
\else\fi
\normalbaselineskip=\smallb@seline pt%
\setbox\strutbox=\hbox{\vrule height.7\normalbaselineskip 
depth.3\baselineskip width0pt}%
\everymath{}%
\def\sc{\csname\smSMALL rm\@nd}\normalbaselines\rm}%
\everydisplay{\indenteddisplay
   \gdef\labeltype{\eqlabel}}%
%
%
\def\hexnumber@#1{\ifcase#1 0\or 1\or 2\or 3\or 4\or 5\or 6\or 7\or 8\or
 9\or A\or B\or C\or D\or E\or F\fi}
\edef\bffam@{\hexnumber@\bffam}
\edef\bdifam@{\hexnumber@\bdifam}
\edef\bsyfam@{\hexnumber@\bsyfam}
\def\undefine#1{\let#1\undefined}
\def\newsymbol#1#2#3#4#5{\let\next@\relax
 \ifnum#2=\thr@@\let\next@\bdifam@\else
 \ifams
 \ifnum#2=\@ne\let\next@\msafam@\else
 \ifnum#2=\tw@\let\next@\msbfam@\fi\fi
 \fi\fi
 \mathchardef#1="#3\next@#4#5}
\def\mathhexbox@#1#2#3{\relax
 \ifmmode\mathpalette{}{\m@th\mathchar"#1#2#3}%
 \else\leavevmode\hbox{$\m@th\mathchar"#1#2#3$}\fi}

\def\bi#1{{\fam\bdifam\relax#1}}
%
%
\ifams\input amsmacro\fi
%
%
\newsymbol\bitGamma 3000
\newsymbol\bitDelta 3001
\newsymbol\bitTheta 3002
\newsymbol\bitLambda 3003
\newsymbol\bitXi 3004
\newsymbol\bitPi 3005
\newsymbol\bitSigma 3006
\newsymbol\bitUpsilon 3007
\newsymbol\bitPhi 3008
\newsymbol\bitPsi 3009
\newsymbol\bitOmega 300A
\newsymbol\balpha 300B
\newsymbol\bbeta 300C
\newsymbol\bgamma 300D
\newsymbol\bdelta 300E
\newsymbol\bepsilon 300F
\newsymbol\bzeta 3010
\newsymbol\bfeta 3011
\newsymbol\btheta 3012
\newsymbol\biota 3013
\newsymbol\bkappa 3014
\newsymbol\blambda 3015
\newsymbol\bmu 3016
\newsymbol\bnu 3017
\newsymbol\bxi 3018
\newsymbol\bpi 3019
\newsymbol\brho 301A
\newsymbol\bsigma 301B
\newsymbol\btau 301C
\newsymbol\bupsilon 301D
\newsymbol\bphi 301E
\newsymbol\bchi 301F
\newsymbol\bpsi 3020
\newsymbol\bomega 3021
\newsymbol\bvarepsilon 3022
\newsymbol\bvartheta 3023
\newsymbol\bvaromega 3024
\newsymbol\bvarrho 3025
\newsymbol\bvarzeta 3026
\newsymbol\bvarphi 3027
\newsymbol\bpartial 3040
\newsymbol\bell 3060
\newsymbol\bimath 307B
\newsymbol\bjmath 307C
\mathchardef\binfty "0\bsyfam@31
\mathchardef\bnabla "0\bsyfam@72
\mathchardef\bdot "2\bsyfam@01
\mathchardef\bGamma "0\bffam@00
\mathchardef\bDelta "0\bffam@01
\mathchardef\bTheta "0\bffam@02
\mathchardef\bLambda "0\bffam@03
\mathchardef\bXi "0\bffam@04
\mathchardef\bPi "0\bffam@05
\mathchardef\bSigma "0\bffam@06
\mathchardef\bUpsilon "0\bffam@07
\mathchardef\bPhi "0\bffam@08
\mathchardef\bPsi "0\bffam@09
\mathchardef\bOmega "0\bffam@0A
\mathchardef\itGamma "0100
\mathchardef\itDelta "0101
\mathchardef\itTheta "0102
\mathchardef\itLambda "0103
\mathchardef\itXi "0104
\mathchardef\itPi "0105
\mathchardef\itSigma "0106
\mathchardef\itUpsilon "0107
\mathchardef\itPhi "0108
\mathchardef\itPsi "0109
\mathchardef\itOmega "010A
\mathchardef\Gamma "0000
\mathchardef\Delta "0001
\mathchardef\Theta "0002
\mathchardef\Lambda "0003
\mathchardef\Xi "0004
\mathchardef\Pi "0005
\mathchardef\Sigma "0006
\mathchardef\Upsilon "0007
\mathchardef\Phi "0008
\mathchardef\Psi "0009
\mathchardef\Omega "000A
%
%
\newcount\firstpage  \firstpage=1  
\newcount\jnl                      
\newcount\secno                    
\newcount\subno                    
\newcount\subsubno                 
\newcount\appno                    
\newcount\tabno                    
\newcount\figno                    
\newcount\countno                  
\newcount\refno                    
\newcount\eqlett     \eqlett=97    
\newif\ifletter
\newif\ifwide
\newif\ifnotfull
\newif\ifaligned
\newif\ifnumbysec  
\newif\ifappendix
\newif\ifnumapp
\newif\ifssf
\newif\ifppt
\newdimen\t@bwidth
\newdimen\c@pwidth
\newdimen\digitwidth                    
\newdimen\argwidth                      
\newdimen\secindent    \secindent=5pc   
\newdimen\textind    \textind=16pt      
\newdimen\tempval                       
\newskip\beforesecskip
\def\beforesecspace{\vskip\beforesecskip\relax}
\newskip\beforesubskip
\def\beforesubspace{\vskip\beforesubskip\relax}
\newskip\beforesubsubskip
\def\beforesubsubspace{\vskip\beforesubsubskip\relax}
\newskip\secskip
\def\secspace{\vskip\secskip\relax}
\newskip\subskip
\def\subspace{\vskip\subskip\relax}
\newskip\insertskip
\def\insertspace{\vskip\insertskip\relax}
\def\sp@ce{\ifx\next*\let\next=\@ssf
               \else\let\next=\@nossf\fi\next}
\def\@ssf#1{\nobreak\secspace\global\ssftrue\nobreak}
\def\@nossf{\nobreak\secspace\nobreak\noindent\ignorespaces}
\def\subsp@ce{\ifx\next*\let\next=\@sssf
               \else\let\next=\@nosssf\fi\next}
\def\@sssf#1{\nobreak\subspace\global\ssftrue\nobreak}
\def\@nosssf{\nobreak\subspace\nobreak\noindent\ignorespaces}
\beforesecskip=24pt plus12pt minus8pt
\beforesubskip=12pt plus6pt minus4pt
\beforesubsubskip=12pt plus6pt minus4pt
\secskip=12pt plus 2pt minus 2pt
\subskip=6pt plus3pt minus2pt
\insertskip=18pt plus6pt minus6pt%
\fontdimen16\tensy=2.7pt
\fontdimen17\tensy=2.7pt
%
%
\def\eqlabel{(\ifappendix\applett
               \ifnumbysec\ifnum\secno>0 \the\secno\fi.\fi
               \else\ifnumbysec\the\secno.\fi\fi\the\countno)}
\def\seclabel{\ifappendix\ifnumapp\else\applett\fi
    \ifnum\secno>0 \the\secno
    \ifnumbysec\ifnum\subno>0.\the\subno\fi\fi\fi
    \else\the\secno\fi\ifnum\subno>0.\the\subno
         \ifnum\subsubno>0.\the\subsubno\fi\fi}
\def\tablabel{\ifappendix\applett\fi\the\tabno}
\def\figlabel{\ifappendix\applett\fi\the\figno}
\def\gac{\global\advance\countno by 1}
%
%

\def\vfootnote#1{\insert\footins\bgroup
\interlinepenalty=\interfootnotelinepenalty
\splittopskip=\ht\strutbox 
\splitmaxdepth=\dp\strutbox \floatingpenalty=20000
\leftskip=0pt \rightskip=0pt \spaceskip=0pt \xspaceskip=0pt%
\noindent\smallfonts\rm #1\ \ignorespaces\footstrut\futurelet\next\fo@t}
%
%
\def\endinsert{\egroup
    \if@mid \dimen@=\ht0 \advance\dimen@ by\dp0
       \advance\dimen@ by12\p@ \advance\dimen@ by\pagetotal
       \ifdim\dimen@>\pagegoal \@midfalse\p@gefalse\fi\fi
    \if@mid \insertspace \box0 \par \ifdim\lastskip<\insertskip
    \removelastskip \penalty-200 \insertspace \fi
    \else\insert\topins{\penalty100
       \splittopskip=0pt \splitmaxdepth=\maxdimen 
       \floatingpenalty=0
       \ifp@ge \dimen@=\dp0
       \vbox to\vsize{\unvbox0 \kern-\dimen@}%
       \else\box0\nobreak\insertspace\fi}\fi\endgroup}   
%
%
%
\def\ind{\hbox to \secindent{\hfill}}
%
%

%
%

%
%
\def\indeqn#1{\alignedfalse\displ@y\halign{\hbox to \displaywidth
    {$\ind\@lign\displaystyle##\hfil$}\crcr #1\crcr}}
%
%
\def\indalign#1{\alignedtrue\displ@y \tabskip=0pt 
  \halign to\displaywidth{\ind$\@lign\displaystyle{##}$\tabskip=0pt
    &$\@lign\displaystyle{{}##}$\hfill\tabskip=\centering
    &\llap{$\@lign\hbox{\rm##}$}\tabskip=0pt\crcr
    #1\crcr}}
\def\indenteddisplay#1$${\indispl@y{#1 }}
\def\indispl@y#1{\disptest#1\eqalignno\eqalignno\disptest}
\def\disptest#1\eqalignno#2\eqalignno#3\disptest{%
    \ifx#3\eqalignno
    \indalign#2%
    \else\indeqn{#1}\fi$$}
%
%

%
%

%
%

%
%

%
%

\def\ns{\noalign{\vskip-3pt}}

%

%
%
\def\bhbar{\rlap{\kern1pt\raise.4ex\hbox{\bf\char'40}}\bi{h}}

\def\frac#1#2{{#1\over#2}}
\ifams
\def\lap{\lesssim}
\def\gap{\gtrsim}

\else

\def\gap{\;\lower3pt\hbox{$\buildrel > \over \sim$}\;}%
\def\lap{\;\lower3pt\hbox{$\buildrel < \over \sim$}\;}\fi

\chardef\ii="10
\def\tqs{\hbox to 25pt{\hfil}}

\def\Bbbone{1\kern-.22em {\rm l}}
%
%
\def\rp{\raise8pt\hbox{$\scriptstyle\prime$}}
%
%
%
%

%
%
\def\[#1\]{\setbox0=\hbox{$\dsty#1$}\argwidth=\wd0
    \setbox0=\hbox{$\left[\box0\right]$}\advance\argwidth by -\wd0
    \left[\kern.3\argwidth\box0\kern.3\argwidth\right]}
%
%
\def\lsb#1\rsb{\setbox0=\hbox{$#1$}\argwidth=\wd0
    \setbox0=\hbox{$\left[\box0\right]$}\advance\argwidth by -\wd0
    \left[\kern.3\argwidth\box0\kern.3\argwidth\right]}
%

%
%

%
\def\pt(#1){({\it #1\/})}
\let\dsty=\displaystyle

%
%
\def\reactions#1{\vskip 12pt plus2pt minus2pt%
\vbox{\hbox{\kern\secindent\vrule\kern12pt%
\vbox{\kern0.5pt\vbox{\hsize=24pc\parindent=0pt\smallfonts\rm NUCLEAR 
REACTIONS\strut\quad #1\strut}\kern0.5pt}\kern12pt\vrule}}}
%
%
\def\slashchar#1{\setbox0=\hbox{$#1$}\dimen0=\wd0%
\setbox1=\hbox{/}\dimen1=\wd1%
\ifdim\dimen0>\dimen1%
\rlap{\hbox to \dimen0{\hfil/\hfil}}#1\else                                        
\rlap{\hbox to \dimen1{\hfil$#1$\hfil}}/\fi}
%
%
\def\textindent#1{\noindent\hbox to \parindent{#1\hss}\ignorespaces}
%
%
\def\opencirc{\raise1pt\hbox{$\scriptstyle{\bigcirc}$}}

\ifams
\def\opensqr{\hbox{$\square$}}

\def\opentridown{\hbox{$\triangledown$}}

\else
\def\opensqr{\vbox{\hrule height.4pt\hbox{\vrule width.4pt height3.5pt
    \kern3.5pt\vrule width.4pt}\hrule height.4pt}}

\def\opentridown{\raise1pt\hbox{$\scriptstyle\bigtriangledown$}}

\fi

%
%
\def\m@th{\mathsurround=0pt}
%
%
\def\cases#1{%
\left\{\,\vcenter{\normalbaselines\openup1\jot\m@th%
     \ialign{$\displaystyle##\hfil$&\rm\tqs##\hfil\crcr#1\crcr}}\right.}%
%
%
\def\oldcases#1{\left\{\,\vcenter{\normalbaselines\m@th
    \ialign{$##\hfil$&\rm\quad##\hfil\crcr#1\crcr}}\right.}
%
%
\def\numcases#1{\left\{\,\vcenter{\baselineskip=15pt\m@th%
     \ialign{$\displaystyle##\hfil$&\rm\tqs##\hfil
     \crcr#1\crcr}}\right.\hfill
     \vcenter{\baselineskip=15pt\m@th%
     \ialign{\rlap{$\phantom{\displaystyle##\hfil}$}\tabskip=0pt&\en
     \rlap{\phantom{##\hfil}}\crcr#1\crcr}}}
\def\ptnumcases#1{\left\{\,\vcenter{\baselineskip=15pt\m@th%
     \ialign{$\displaystyle##\hfil$&\rm\tqs##\hfil
     \crcr#1\crcr}}\right.\hfill
     \vcenter{\baselineskip=15pt\m@th%
     \ialign{\rlap{$\phantom{\displaystyle##\hfil}$}\tabskip=0pt&\enpt
     \rlap{\phantom{##\hfil}}\crcr#1\crcr}}\global\eqlett=97
     \global\advance\countno by 1}
%
%
\def\eq(#1){\ifaligned\@mp(#1)\else\hfill\llap{{\rm (#1)}}\fi}
\def\ceq(#1){\ns\ns\ifaligned\@mp\fi\eq(#1)\cr\ns\ns}
\def\eqpt(#1#2){\ifaligned\@mp(#1{\it #2\/})
                    \else\hfill\llap{{\rm (#1{\it #2\/})}}\fi}
\let\eqno=\eq
%
%
\countno=1

\def\aleq{&\rm(\ifappendix\applett
               \ifnumbysec\ifnum\secno>0 \the\secno\fi.\fi
               \else\ifnumbysec\the\secno.\fi\fi\the\countno}
\def\noaleq{\hfill\llap\bgroup\rm(\ifappendix\applett
               \ifnumbysec\ifnum\secno>0 \the\secno\fi.\fi
               \else\ifnumbysec\the\secno.\fi\fi\the\countno}
\def\@mp{&}
\def\en{\ifaligned\aleq)\else\noaleq)\egroup\fi\gac}
\def\cen{\ns\ns\ifaligned\@mp\fi\en\cr\ns\ns}
\def\enpt{\ifaligned\aleq{\it\char\the\eqlett})\else
    \noaleq{\it\char\the\eqlett})\egroup\fi
    \global\advance\eqlett by 1}
\def\endpt{\ifaligned\aleq{\it\char\the\eqlett})\else
    \noaleq{\it\char\the\eqlett})\egroup\fi
    \global\eqlett=97\gac}
%
%

\def\JPA{{\it J. Phys. A: Math. Gen.}}



%
%

\def\PRL{{\it Phys. Rev. Lett.}}

\headline={\ifodd\pageno{\ifnum\pageno=\firstpage\hfill
   \else\rrhead\fi}\else\lrhead\fi}
\def\rrhead{\textfonts\hskip\secindent\it
    \shorttitle\hfill\rm\folio}
\def\lrhead{\textfonts\hbox to\secindent{\rm\folio\hss}%
    \it\aunames\hss}
\footline={\ifnum\pageno=\firstpage \hfill\textfonts\rm\folio\fi}
\def\@rticle#1#2{\vglue.5pc
    {\parindent=\secindent \bf #1\par}
     \vskip2.5pc
    {\exhyphenpenalty=10000\hyphenpenalty=10000
     \baselineskip=18pt\raggedright\noindent
     \headfonts\bf#2\par}\futurelet\next\sh@rttitle}%
\def\title#1{\gdef\shorttitle{#1}
    \vglue4pc{\exhyphenpenalty=10000\hyphenpenalty=10000 
    \baselineskip=18pt 
    \raggedright\parindent=0pt
    \headfonts\bf#1\par}\futurelet\next\sh@rttitle} 

\def\article#1#2{\gdef\shorttitle{#2}\@rticle{#1}{#2}} 
\def\review#1{\gdef\shorttitle{#1}%
    \@rticle{REVIEW \ifpbm\else ARTICLE\fi}{#1}}
\def\topical#1{\gdef\shorttitle{#1}%
    \@rticle{TOPICAL REVIEW}{#1}}
\def\comment#1{\gdef\shorttitle{#1}%
    \@rticle{COMMENT}{#1}}
\def\note#1{\gdef\shorttitle{#1}%
    \@rticle{NOTE}{#1}}
\def\prelim#1{\gdef\shorttitle{#1}%
    \@rticle{PRELIMINARY COMMUNICATION}{#1}}
\def\letter#1{\gdef\shorttitle{Letter to the Editor}%
     \gdef\aunames{Letter to the Editor}
     \global\lettertrue\ifnum\jnl=7\global\letterfalse\fi
     \@rticle{LETTER TO THE EDITOR}{#1}}
\def\sh@rttitle{\ifx\next[\let\next=\sh@rt
                \else\let\next=\f@ll\fi\next}
\def\sh@rt[#1]{\gdef\shorttitle{#1}}
\def\f@ll{}
\def\author#1{\ifletter\else\gdef\aunames{#1}\fi\vskip1.5pc
    {\parindent=\secindent  
     \hang\textfonts  
     \ifppt\bf\else\rm\fi#1\par}  
     \ifppt\bigskip\else\smallskip\fi
     \futurelet\next\@unames}
\def\@unames{\ifx\next[\let\next=\short@uthor
                 \else\let\next=\@uthor\fi\next}
\def\short@uthor[#1]{\gdef\aunames{#1}}
\def\@uthor{}
\def\address#1{{\parindent=\secindent
    \exhyphenpenalty=10000\hyphenpenalty=10000
\ifppt\textfonts\else\smallfonts\fi\hang\raggedright\rm#1\par}%
    \ifppt\bigskip\fi}
\def\jl#1{\global\jnl=#1}
\jl{0}%
\def\journal{\ifnum\jnl=1 J. Phys.\ A: Math.\ Gen.\ 
        \else\ifnum\jnl=2 J. Phys.\ B: At.\ Mol.\ Opt.\ Phys.\ 
        \else\ifnum\jnl=3 J. Phys.:\ Condens. Matter\ 
        \else\ifnum\jnl=4 J. Phys.\ G: Nucl.\ Part.\ Phys.\ 
        \else\ifnum\jnl=5 Inverse Problems\ 
        \else\ifnum\jnl=6 Class. Quantum Grav.\ 
        \else\ifnum\jnl=7 Network\ 
        \else\ifnum\jnl=8 Nonlinearity\
        \else\ifnum\jnl=9 Quantum Opt.\
        \else\ifnum\jnl=10 Waves in Random Media\
        \else\ifnum\jnl=11 Pure Appl. Opt.\ 
        \else\ifnum\jnl=12 Phys. Med. Biol.\
        \else\ifnum\jnl=13 Modelling Simulation Mater.\ Sci.\ Eng.\ 
        \else\ifnum\jnl=14 Plasma Phys. Control. Fusion\ 
        \else\ifnum\jnl=15 Physiol. Meas.\ 
        \else\ifnum\jnl=16 Sov.\ Lightwave Commun.\
        \else\ifnum\jnl=17 J. Phys.\ D: Appl.\ Phys.\
        \else\ifnum\jnl=18 Supercond.\ Sci.\ Technol.\
        \else\ifnum\jnl=19 Semicond.\ Sci.\ Technol.\
        \else\ifnum\jnl=20 Nanotechnology\
        \else\ifnum\jnl=21 Meas.\ Sci.\ Technol.\ 
        \else\ifnum\jnl=22 Plasma Sources Sci.\ Technol.\ 
        \else\ifnum\jnl=23 Smart Mater.\ Struct.\ 
        \else\ifnum\jnl=24 J.\ Micromech.\ Microeng.\
   \else Institute of Physics Publishing\ 
   \fi\fi\fi\fi\fi\fi\fi\fi\fi\fi\fi\fi\fi\fi\fi
   \fi\fi\fi\fi\fi\fi\fi\fi\fi}
\let\abs=\beginabstract

\let\endabs=\endabstract
\def\submitted{\ifppt\noindent\textfonts\rm Submitted to \journal\par
     \bigskip\fi}
\def\today{\number\day\ \ifcase\month\or
     January\or February\or March\or April\or May\or June\or
     July\or August\or September\or October\or November\or
     December\fi\space \number\year}
\def\date{\ifppt\noindent\textfonts\rm 
     Date: \today\par\goodbreak\bigskip\fi}
%
%
\def\pacs#1{\ifppt\noindent\textfonts\rm 
     PACS number(s): #1\par\bigskip\fi}
%

%
%
\def\section#1{\ifppt\ifnum\secno=0\eject\fi\fi
    \subno=0\subsubno=0\global\advance\secno by 1
    \gdef\labeltype{\seclabel}\ifnumbysec\countno=1\fi
    \goodbreak\beforesecspace\nobreak
    \noindent{\bf \the\secno. #1}\par\futurelet\next\sp@ce}
\def\subsection#1{\subsubno=0\global\advance\subno by 1
     \gdef\labeltype{\seclabel}%
     \ifssf\else\goodbreak\beforesubspace\fi
     \global\ssffalse\nobreak
     \noindent{\it \the\secno.\the\subno. #1\par}%
     \futurelet\next\subsp@ce}
\def\subsubsection#1{\global\advance\subsubno by 1
     \gdef\labeltype{\seclabel}%
     \ifssf\else\goodbreak\beforesubsubspace\fi
     \global\ssffalse\nobreak
     \noindent{\it \the\secno.\the\subno.\the\subsubno. #1}\null. 
     \ignorespaces}
%

%
%
\def\numappendix#1{\ifappendix\ifnumbysec\countno=1\fi\else
    \countno=1\figno=0\tabno=0\fi
    \subno=0\global\advance\appno by 1
    \secno=\appno\gdef\applett{A}\gdef\labeltype{\seclabel}%
    \global\appendixtrue\global\numapptrue
    \goodbreak\beforesecspace\nobreak
    \noindent{\bf Appendix \the\appno. #1\par}%
    \futurelet\next\sp@ce}
\def\numsubappendix#1{\global\advance\subno by 1\subsubno=0
    \gdef\labeltype{\seclabel}%
    \ifssf\else\goodbreak\beforesubspace\fi
    \global\ssffalse\nobreak
    \noindent{\it A\the\appno.\the\subno. #1\par}%
    \futurelet\next\subsp@ce}
\def\@ppendix#1#2#3{\countno=1\subno=0\subsubno=0\secno=0\figno=0\tabno=0
    \gdef\applett{#1}\gdef\labeltype{\seclabel}\global\appendixtrue
    \goodbreak\beforesecspace\nobreak
    \noindent{\bf Appendix#2#3\par}\futurelet\next\sp@ce}
\def\Appendix#1{\@ppendix{A}{. }{#1}}
\def\appendix#1#2{\@ppendix{#1}{ #1. }{#2}}
\def\App#1{\@ppendix{A}{ }{#1}}
\def\app{\@ppendix{A}{}{}}
\def\subappendix#1#2{\global\advance\subno by 1\subsubno=0
    \gdef\labeltype{\seclabel}%
    \ifssf\else\goodbreak\beforesubspace\fi
    \global\ssffalse\nobreak
    \noindent{\it #1\the\subno. #2\par}%
    \nobreak\subspace\noindent\ignorespaces}
%
%
\def\@ck#1{\ifletter\bigskip\noindent\ignorespaces\else
    \goodbreak\beforesecspace\nobreak
    \noindent{\bf Acknowledgment#1\par}%
    \nobreak\secspace\noindent\ignorespaces\fi}
\def\ack{\@ck{s}}
\def\ackn{\@ck{}}
\def\n@ip#1{\goodbreak\beforesecspace\nobreak
    \noindent\smallfonts{\it #1}. \rm\ignorespaces}
\def\naip{\n@ip{Note added in proof}}
\def\na{\n@ip{Note added}}

%
%

%

%
%

%

%
\def\table#1{\tablecaption{#1}}
\def\tablecont{\topinsert\global\advance\tabno by -1
    \tablecaption{(continued)}}
\def\tablecaption#1{\gdef\labeltype{\tablabel}\global\widefalse
    \leftskip=\secindent\parindent=0pt
    \global\advance\tabno by 1
    \smallfonts{\bf Table \ifappendix\applett\fi\the\tabno.} \rm #1\par
    \smallskip\futurelet\next\t@b}
\def\t@b{\ifx\next*\let\next=\widet@b
             \else\ifx\next[\let\next=\fullwidet@b
                      \else\let\next=\narrowt@b\fi\fi
             \next}
\def\widet@b#1{\global\widetrue\global\notfulltrue
    \t@bwidth=\hsize\advance\t@bwidth by -\secindent} 
\def\fullwidet@b[#1]{\global\widetrue\global\notfullfalse
    \leftskip=0pt\t@bwidth=\hsize}                  
\def\narrowt@b{\global\notfulltrue}
\def\align{\catcode`?=13\ifnotfull\moveright\secindent\fi
    \vbox\bgroup\halign\ifwide to \t@bwidth\fi
    \bgroup\strut\tabskip=1.2pc plus1pc minus.5pc}
\def\endalign{\egroup\egroup\catcode`?=12}

%
%
\def\L#1{#1\hfill}

%
%
\def\br{\noalign{\vskip2pt\hrule height1pt\vskip2pt}}
\def\mr{\noalign{\vskip2pt\hrule\vskip2pt}}
%

%
%

%

\catcode`?=13
\def\lineup{\setbox0=\hbox{\smallfonts\rm 0}%
    \digitwidth=\wd0%
    \def?{\kern\digitwidth}%
    \def\\{\hbox{$\phantom{-}$}}%
    \def\-{\llap{$-$}}}
\catcode`?=12
%
%
\def\sidetable#1#2{\hbox{\ifppt\hsize=18pc\t@bwidth=18pc
                          \else\hsize=15pc\t@bwidth=15pc\fi
    \parindent=0pt\vtop{\null #1\par}%
    \ifppt\hskip1.2pc\else\hskip1pc\fi
    \vtop{\null #2\par}}} 
\def\lstable#1#2{\everypar{}\tempval=\hsize\hsize=\vsize
    \vsize=\tempval\hoffset=-3pc
    \global\tabno=#1\gdef\labeltype{\tablabel}%
    \noindent\smallfonts{\bf Table \ifappendix\applett\fi
    \the\tabno.} \rm #2\par
    \smallskip\futurelet\next\t@b}
\def\inctabno{\global\advance\tabno by 1}
%
%
\def\Figures{\vfill\eject\global\appendixfalse\textfonts\rm
    \everypar{}\noindent{\bf Figure captions}\par
    \bigskip}
\def\figure#1{\figc@ption{#1}\bigskip}
\def\figc@ption#1{\global\advance\figno by 1\gdef\labeltype{\figlabel}%
   {\parindent=\secindent\smallfonts\hang
    {\bf Figure \ifappendix\applett\fi\the\figno.} \rm #1\par}}
%
%
\def\refHEAD{\goodbreak\beforesecspace
     \noindent\textfonts{\bf References}\par
     \let\ref=\rf
     \nobreak\smallfonts\rm}
\def\references{\refHEAD\parindent=0pt
     \everypar{\hangindent=18pt\hangafter=1
     \frenchspacing\rm}%
     \secspace}
\def\rf#1{\par\noindent\hbox to 21pt{\hss #1\quad}\ignorespaces}
%

%

%
%
\def\numrefjl#1#2#3#4#5{\par\rf{#1}#2 {\it #3 \bf #4} #5\par}
%
%
\def\numrefbk#1#2#3#4{\par\rf{#1}#2 {\it #3} #4\par}
%
%

%
%

%
\catcode`\@=12
%
%
\def\pptstyle{\ppttrue\headsize{17}{24}%
\textsize{12}{16}%
\smallsize{10}{12}%
\hsize=37.2pc\vsize=56pc
\textind=20pt\secindent=6pc}
%
%

%
%

%
%

%
\parindent=\textind

\pptstyle
\jl{1}

\title{Determination of the bond percolation threshold for the Kagom\'e lattice}[Bond
percolation threshold on the Kagom\'e lattice]

\author{Robert M. Ziff and Paul N. Suding}

\address{Department of Chemical Engineering, University of Michigan,
Ann Arbor MI 48109-2136 USA}

\abs
The hull-gradient method is used to determine the critical threshold
for bond percolation on the two-dimensional Kagom\'e lattice (and its dual, the dice
lattice).
For this system, the hull walk is represented as
a self-avoiding trail, or mirror-model trajectory, on
the (3,4,6,4)-Archimedean tiling lattice.  The result
$p_c = 0.524\,405\,3 \pm 0.000\,000\,3$ (one standard deviation of error)
is  not consistent with previously conjectured values.
\endabs

\submitted

\date 

\pacs{64.60.Ak}

\section{Introduction}  

The Kagom\'e lattice (Fig.~1) is one of the fundamental lattices
of two-dimensional percolation, as well as many other
two-dimensional lattice problems.
It is one of the eleven Archimedean tiling lattices (in 
which all vertices are of the same type), designated as
(3,6,3,6) in the notation of [1],
which means that each vertex touches a triangle,
hexagon, triangle, and hexagon.
\ The Kagom\'e lattice is
intimately related to other important lattices in percolation;
its sites 
correspond to the bonds of the  honeycomb lattice, which implies that
the percolation thresholds $p_c$ for those two systems are the same,
$1 - 2 \sin (\pi/18)$, and by duality they are also
equal to one minus the threshold for bond percolation
on the triangular lattice [2].

For bond percolation on the Kagom\'e lattice,
no exact expression for
$p_c$  is known, although two conjectures were made
 a number of years ago, both
 within the larger context of the $q$-state
Potts model from which percolation follows in the limit of $q = 1$.
Wu [3] conjectured that $p_c$ is the solution to
$$p^6 - 6p^5 + 12p^4 - 6p^3 - 3p^2 + 1 = 0 \eqno(1)$$
which yields 
$$p_c(\hbox{Wu}) = 0.524\,429\,718  \ . \eqno(2)$$
Subsequently, Enting and Wu [4] showed that the
general conjecture is not valid  for $q = 3$, thereby
casting doubt on its validity for $q=1$.
Tsallis [5] conjectured   that $p_c$ satisfies
$$-p^3 + p^2 + p = 1 - 2 \sin(\pi/18)   \eqno(3)$$
which yields 
$$p_c(\hbox{Tsallis}) = 0.522\,372\,078 \ .   \eqno(4)$$
Note that the quantity on the right-hand side of (3)  is identical to the threshold
for site percolation on the Kagom\'e lattice and is the solution
to the cubic equation $y^3 - 3 y^2 + 1 = 0$, so it follows that (4) is the solution
to 
the ninth order equation
$$p^9 - 3 p^8 + 8 p^6 - 6 p^5 - 6 p^4 + 5 p^3  + 3 p^2 - 1 = 0  \ . \eqno(5)$$

The only existing numerical values of $p_c$ of relatively
high precision appear to be those of Yonezawa et al.\ [6],
who find
 $0.5244 \pm 0.0002$, and van der Marck [7],
who finds
$0.5243 \pm 0.0004$.
These results clearly favor
Wu's value over Tsallis'.  Note that
Hu, Chen and Wu [8] have also recently presented numerical evidence that
Wu's conjecture still
works quite well (and better than Tsallis') for the 
Potts model of various $q$.
In order to investigate the validity of these conjectures further,
and to provide an accurate value of $p_c$ for use by others [9],
we have carried out a new numerical study to determine the percolation
threshold for bond percolation on the Kagom\'e lattice.

\section{Method}  

The method we employed is the hull-gradient method [10],
in which the gradient-percolation
frontier is created by a hull-generating walk.
In gradient percolation [11,12],
a linear gradient in $p$ is imposed on the lattice in the vertical 
direction; as the height increases, the occupied bond density
$p$ also increases.  The estimate of $p_c$ is related to the average
position of the frontier of the percolating region.
To simultaneously create and measure that frontier,
a hull-generating
walk is employed [13,14].  In this walk, the status of a bond
(whether occupied or vacant) is determined when the bond is 
visited by generating a random number and comparing that number to
the occupation probability for that height.

The efficiency of this method derives from the fact that in the hull-generating
method, the entire lattice is not filled before the walk begins.
Rather, the lattice is initialized with all bonds undetermined, and 
the state of the bonds are decided only when they are visited.
If the walk does not reach a given bond, then the status 
of that bond remains undetermined, and no random number need be 
generated.  The error of this method is
at the statistical limit 
$ 0.5 N^{-1/2}$, where $N$
is the total quantity of random numbers generated.
Previously, we used the hull-gradient method
to find $p_c$  for site percolation on the square lattice
to six significant figures, $0.592\,746\,0 \pm 0.000\,000\,5$ [10,15].
This value was confirmed using a different method
(also implemented using hulls) [16],
and is consistent with the most precise value  $0.592\,77 \pm 0.000\,05$
[17] obtained using the more traditional average crossing-probability method [18].
In [10], we also applied the hull-gradient method to determine
$p_c$ for site percolation on the Kagom\'e
lattice, and found a value
($0.652\,704 \pm 0.000\,009$) in agreement with the theoretical
prediction mentioned above.

For bond percolation, the hull walk simplifies to a trajectory that ``bounces"
back and forth between the centers of the occupied and vacant bonds of the hull, as 
first noted by Grassberger [19] for the case of bond percolation on the square lattice.
For the Kagom\'e lattice, a similar method can be used.
As shown in Fig.~1, the walk moves along line segments that connect centers
of adjacent bonds.  These line segments produce a new lattice whose topology is
the Archimedean (6,4,3,4)-lattice [1] shown in Fig.~2. \  Here,
the walker turns clockwise when an occupied bond is hit, and
counter-clockwise when a vacant bond is hit, so that the bonds 
are effectively rotators [20] or mirrors [21,22]
on the vertices of the (6,4,3,4)-lattice.
An occupied bond (probability $ p$) 
on the Kagom\'e lattice corresponds to a mirror placed 
tangent to the vertex of the hexagon on the (6,4,3,4) lattice,
while a vacant bond (probability $1-p$)
corresponds to a mirror that intersects the hexagon.
The hull walk is then a mirror-model trajectory [21,22] on the
(6,4,3,4)-lattice.

Many other representations of the hull walk can also be made.
The vacant bonds on the Kagom\'e lattice can be associated with occupied
bonds on the dice (or ``diced") lattice shown in Fig.~3, which is dual to the
Kagom\'e lattice, and the walk creates a hull on that lattice also.
The hull trajectory is also a self-avoiding
trail on the directed (6,4,3,4)-lattice, with opposing direction
vectors at each vertex.  The hulls on this lattice can also
be produced by a random tiling similar to [23], with ``kite"-shaped
tiles having weights $p$ and $1-p$ as shown in Fig.~4.

In gradient percolation, the hull of the percolating region
resides on bonds whose average value of $p$ gives an
estimate $p_c(g)$, which approaches $p_c$ as the gradient $g \equiv |\nabla p|$ goes to
zero [11,12].
Equivalent to taking the average value of $p$ on bonds of the 
hull, one can simply take as the estimate [12]
$$ p_c(g) = { n_{\hbox{occ}} \over n_{\hbox{occ}} + n_{\hbox{vac}}} \ ,  \eqno(6) $$
since
the expected fraction of occupied bonds in the hull equals the average value of $p$
of the vertices that belong to the hull. 
Here, $n_{\hbox{occ}}$ is the number of vertices corresponding to
occupied bonds and $n_{\hbox{vac}}$ is the number of vertices corresponding to
vacant bonds in the hull.

To represent the (6,4,3,4)-lattice in the computer efficiently,
it is necessary
transform it to align on a rectilinear grid.
One way to do this is shown in Fig.~5, where the basic
rectangle of six sites is repeated in every second column and every third row.
While we have distorted the lattice laterally
to accommodate the square lattice periodicity,
we have not shifted any vertices vertically
with respect to each other, so as not to affect the gradient
in the vertical direction.
In producing that gradient, we made the change
in $p$ proportional to the actual height,
so that changes between the wider rows in Fig.~2
equal twice the change of the narrower ones.  The 
gradient $g$ is defined here by 
$g = \Delta p / \ell $
where $\Delta p$ is the change of $p$ between the wider rows,
and $\ell$ is the bond length, taken to be unity.
(We also considered two alternate representations where the gradient was not 
precisely uniform on a local scale; the behavior of these systems is
discussed in the Appendix.)
A $3d$  array was used to store the six possible outgoing directions based upon the 
two incoming directions, the six types of vertices, and the status
(occupied or vacant) of the vertex.

The lattice was initialized by filling the first column
halfway with occupied bonds
and the rest with vacant bonds, which prevents
the walk from closing on itself at the start.
With the gradient in the vertical direction, the 
walk naturally drifts to the right.
Periodic boundary conditions were applied in the horizontal direction,
and  each new column to the right was cleared off as it was first visited.
This allows the simulation to run indefinitely and have
essentially no boundary effects from the horizontal ends of the system.
We tracked the maximum distance
the walk wandered to the left of the moving front in order
to confirm that the system width was
sufficient to preclude wraparound errors. 

To rule out systematic errors related to random number generation,
three different generators were tried.
For most of the runs, we used 
the shift-register sequence generator R$_7$(9689)
[16, 24] defined by 
$$ x_n = x_{n-471} \ ^\wedge x_{n-1586} \ ^\wedge x_{n-6988} \ ^\wedge x_{n-9689} \eqno(7)$$
where $\ ^\wedge$ is the bitwise exclusive-or operation.
This ``four-tap" generator is equivalent to  decimating by 7 (taking every 
seventh term) of the sequence
generated by the two-tap rule  R(9689), $ x_n = x_{n-471} \ ^\wedge x_{n-9689}$,
which follows from [25].
This decimation
has the effect
of vastly reducing the three- and four-point correlations of the two-tap
generator, and was previously found to yield good behavior
for problems of this type [26].  For the system with the second
smallest gradient, we considered
two additional random number generators. The second generator, R$_{21}$(9689), 
was obtained by further decimating R$_{7}$(9689) three times,
simply by using every third number,
which may yield better statistical properties.
For the third generator, we used a traditional congruential generator CONG,
but with a very large modulus of 64 bits [27]:
$$ x_n = (5\,081\,641\,266\,417\,562\,522 x_{n-1} + 11) \hbox{ mod } 2^{64} \ .  \eqno(8)  $$

First, to study the general finite-size behavior, we considered lattices
of height $H = 128, 256, 512, 1024, 2048$ and 4096, with widths
sufficient to avoid wraparound error, and $p$ ranging all the way
from 0 to 1.  For these systems, $g = 1.5/H$, the factor of 1.5
resulting from the changing increment of $p$ between the  wide and narrow
rows as described above.  
Approximately $10^{11}$ occupied plus vacant
vertices were generated for each of these
lattices.  Then, to obtain a precise final value, we used systems with  very
small gradients $g = 2.564 \cdot 10^{-5} $ and $7.324 \cdot 10^{-6} $ by using
lattices of height 4096 and 8192, with $p$ ranging from
$0.49$ to $0.56$ and $0.505$ to $0.545$, respectively;
$2 \cdot 10^{12}$ steps were carried out for each of these systems.
In all, several months of workstation
computer time were used to obtain the final results.

In the simulations, we kept track of the maximum and minimum
heights of the hull.  As $g$ decreases,
the relative width of the walk decreases
as $g^{3/7}$ [12], allowing us to expand the gradient
as mentioned above.  Note that we also carried out runs on a system
of height 64, but  ran into difficulty because the walk wandered all
the way to a top or bottom boundary where it got stuck in a dead end.
Presumably, this could be
averted by more carefully constructing those boundaries,
but we did not attempt to do it.

\section{Results and Discussion}

First we compare the three random number generators.
Table 1 gives the results for runs for $g = 2.564 \cdot 10^{-5}$,
where all generators were used. 
Error bars represent one standard deviation, and follow from the statistical
formula $ [p_c(1-p_c)/N]^{1/2} \approx 0.5 N^{-1/2}$
where $N=n_{\rm occ} + n_{\rm vac}$,  since the
occupancy of each vertex is achieved with complete statistical independence.
Clearly, the three random number generators give statistically consistent
results, and we thus
averaged their results to get the final data point for this $g$.

Our complete results are shown in Fig.~6, where we plot
$p_c(g)$ vs.\ $g$ for all the lattices we considered.  
This plot provides good evidence that the dependence of $p_c(g)$ upon $g$
is linear (as observed by Rosso et al.\ [12]
for the case of site percolation on the square lattice) with the behavior
$$ p_c(g) =p_c + 0.010\, g \eqno(9) $$
The slope $0.010$ implies that the average
value of $p$ differs from $p_c$ by an amount corresponding to
one hundredth of a lattice spacing, independent of the gradient. 
For $g = 2.564 \cdot 10^{-5}$, this implies a finite-size correction
$p_c(g) -  p_c = 2.3 \cdot 10^{-7}$
that is somewhat smaller than the error bars given in 
Table 1.  For 
 $g = 7.324 \cdot 10^{-6}$, 
the simulations of $2.2 \cdot 10^{12}$ steps using R$_7$(9689)
yielded  $0.524\,405\,5 \pm 3.4 \cdot 10^{-7}$.
Here, the finite-size correction is insignificant
compared to the statistical error.
Putting these results all together, we obtain our final result
$$ p_c = 0.524\,405\,3 \pm 0.000\,000\,3  \eqno(10) $$
This result is consistent with --- but nearly 1000 times more precise 
than --- previous values [6,7].
It evidently agrees with neither Tsallis' nor Wu's predictions,
although the difference with Wu's approximate conjecture
is remarkably small, only 47 parts per million (but still much
larger than our error bars of less than one part per million).  Note that it would
be quite difficult to observe this small difference in $p_c$ using conventional
methods [e.g., 6, 7, 8, 17, 18, 28, 29]. 

If neither Wu's nor Tsallis' conjecture is valid, is there
perhaps some other simple polynomial that yields
$p_c$?
In the absence of a theory, we can search numerically for possible candidates 
consistent with our numerical result.  However, if we allow the maximum order of the
polynomial to be
six, and the integer coefficients to be as large as say $\pm 24$
(except for the leading coefficient, which we restrict to unity),
then we find literally thousands of  polynomials with roots
within two standard deviations of (10).
Some examples are:
$$ p^4 + 7 p^3 + 17 p - 10 = 0  \ ,  \quad p_c = 0.524\,405\,335 \eqno(11a)$$
$$ p^4 - 24 p^2 +  p + 6 = 0 \ ,   \quad p_c = 0.524\,405\,671 \eqno(11b)$$
$$ p^5 - 2 p^4 - 2 p^3 + 16 p^2 - 4 = 0  \ ,  \quad p_c = 0.524\,405\,424 \eqno(11c) $$
$$ p^5 + 5 p^3 - 8  p^2 +18  p - 8 = 0  \ ,  \quad p_c = 0.524\,404\,863 \eqno(11d) $$
$$ p^6 + 3 p^5  -   3 p^3 + 12  p - 6 = 0 \ ,  \quad p_c = 0.524\,405\,290  \eqno(11e)$$
$$ p^6 + 5 p^5 + 7 p^3 - 5  p^2 + 6  p - 3 = 0 \ ,  \quad p_c = 0.524\,405\,306  \eqno(11f)$$
$$ p^6 + 3 p^5  + 9 p^4 -   p^3 +   p^2 + 2  p - 2 = 0 \ ,  \quad p_c = 0.524\,405\,134  \eqno(11g)$$
Note also that $(11/40)^{1/2} = 0.524\,404\,424$ is only slightly low.
Unfortunately, 
it is not possible by numerical means
to determine $p_c$ with sufficient accuracy
to distinguish which of these many polynomials
is the correct  one (if indeed one is!).

Note finally that (10) implies that the bond threshold of the dice lattice 
shown in Fig.~3 is given by
$$ p_c(\hbox{dice}) = 1 - p_c(\hbox{Kagom\'e}) = 
0.475\,594\,7 \pm 0.000\,000\,3  \eqno(12) $$

\vfill\eject
\topinsert
\table{Results for $p_c(g)$ given by (6) for
runs with height $H = 4096$ and
 gradient  $g = 0.000\, 025\, 64$, using
three different random number generators (RNG).  $N$ is the
total number of occupied and vacant bonds generated, and $\sigma$
represents one standard deviation of error (68\% confidence interval).}
\align\L{#}&\L{$#$}&\L{$#$}&\L{$#$}\cr
\br
RNG&N&p_c(g)&\sigma= 0.5 N^{-0.5}\cr
\mr
R$_7(9689)$&1.0\cdot 10^{12}&0.524\,404\,8&\pm 0.000\,000\,5\cr
R$_{21}(9689)$&0.5\cdot 10^{12}&0.524\,405\,3&\pm 0.000\,000\,7\cr
CONG(64-bit)&0.5\cdot 10^{12}&0.524\,405\,9&\pm 0.000\,000\,7\cr
Average&2.0\cdot 10^{12}&0.524\,405\,2&\pm  0.000\,000\,4\cr
\br
\endalign
\endinsert
\null\vfill\eject

\Appendix{}

Besides the system described above with the gradient applied completely
uniformly, we also considered two systems in which
gradient was not constructed so precisely on a local scale, and it is instructive
to see their effects on the finite-size behavior.
At first, we squared-off the (6,3,4,3)-lattice by
``stretching" it horizontally, leading to a lattice similar to Fig.~5 but rotated
by 90$^\circ$.  
Thus, we effectively pushed up and down alternating columns in the 
original lattice, and
the gradient was applied equally between all the rows in this distorted lattice.
The idea was that these local variations should have little effect on the
behavior when the gradient is small.  However, the deviations turned out to be
rather large, until the gradient dropped to about $0.002$, as shown 
in Fig.~7 (case A).
As a second test (case B in Fig.~7), we represented the lattice as in Fig.~5, but
applied the gradient equally between all rows, whether ``wide" or ``narrow."  
Again, the behavior of the finite-size corrections to $p_c(g)$ differed from the
uniform case, but not as much here.
In the limit of $g$ small, where $p$ changes very little from row to row,
all systems followed the same limiting finite-size behavior as (9).
 These results
show, however, that for the linear behavior to remain valid
for moderately large gradients, a 
uniform gradient must be applied to the lattice in its actual configuration.

\ack
        
The authors thank A J Guttmann for interesting them in this problem,
and thank F. Y. Wu for comments.                                   
This material is based upon work supported by the U. S. 
National Science Foundation under Grant No.\ DMR-9520700. 
                     
\references

\def \j #1;#2;#3;#4;#5;#6;{\numrefjl{[#1]}{#2 #6}{#3}{#4}{#5}}
\def \b #1;#2;#3;#4;#5;{\numrefbk{[#1]}{#2 #5}{#3}{#4}}
  
E-mail addresses: rziff@umich.edu, psuding@umich.edu

\b 1; Gr\" unbaum B and Shephard G C;Tilings and Patterns;(New York: Freeman and Co.);1987;

\j 2; Sykes M F and Essam J W;J. Math. Phys.;5;1117;1964;

\j 3;Wu F Y;J. Phys. C: Solid State;12;L645;1979;

\j 4;Enting I G and Wu F Y;J. Stat. Phys.;28;351;1982;

\j 5;Tsallis C;J. Phys. C: Solid State;15;L757;1982;

\j 6;Yonezawa F, Sakamoto S and Hori M;Phys. Rev. B;40;636;1989;

\j 7;van der Marck S C;Phys. Rev. E;55;1514;1997;

\j 8;Hu C-K, Chen J-A, and Wu F Y;Mod. Phys. Lett. B;8;455;1994;

\j 9; Guttmann A J and Jensen I;;;(to be published);1997;

\j 10;Ziff R M and Sapoval B;\JPA;18;L1169;1986;

\j 11;Sapoval B, Rosso M and Gouyet J F;J. Physique Lett. (Paris);46;L149;1985;

\j 12;Rosso M, Gouyet J F and Sapoval B;Phys. Rev. B;32;6053;1986;

\j 13;Ziff R M, Cummings P T and Stell G;\JPA;17;3009;1984;

\j 14;Weinrib A and Trugman S;Phys. Rev. B;31;2993;1985;

\j 15;Ziff R M and Stell G;UM LaSC Report 88-4;;(unpublished);1988;

\j 16;Ziff R M;\PRL;69;2670;1992;

\j 17;Gebele T;\JPA;17;L51;1984;

\b 18;Stauffer D and Aharony A;An Introduction to Percolation Theory, 2nd revised
edition; (London: Taylor and Francis);1994;

\j 19;Grassberger P;\JPA;19;2675;1984;

\j 20;Gunn J M F and Ortu\~no M;\JPA;18;L1096;1985;

\j 21;Cao M-S and Cohen E G D;J. Statis. Phys.;;(to be published);1997;

\j 22;Wang F  and Cohen E G D;J. Statis. Phys.;81;467;1995;

\j 23;Roux S, Guyon E and Sornette D;\JPA;21;L475;1988;

\b 24;Golomb S W;Shift Register Sequences;(San Francisco: Holden-Day);1967;

\j 25;Zierler N;Inform. Control;15;67;1969;

\j 26;Ziff R M;unpublished;;;;

\j 27;Wood W W;(Los Alamos, NM);;private communication;;

\j 28;Reynolds P J, Stanley H E and Klein W;Phys. Rev. B;21;1223;1980;

\j 29;Hu C-K; Phys. Rev. Lett.;69;2739;1992;



\Figures

\figure{The Kagom\'e lattice, showing a path of bonds that represent
the occupied bonds of the frontier of the percolating region,
which is above it (gradient increases in the vertical direction).
The dashed line shows the hull walk, which ``bounces" back and forth
between centers of the occupied and vacant bonds of the hull.}

\figure{The (6,4,3,4)-Archimedean lattice, on which the hull walk of Fig.~2
effectively takes place.  The trajectory of the walk is equivalent to a
mirror-model trajectory [21,22] on the  (6,4,3,4)-lattice,
in which the bonds on the underlying
Kagom\'e lattice act as the mirrors.}

\figure{The same hull as in Fig.~2 on the dice (or ``diced") lattice, the dual to the
Kagom\'e lattice. Vacant bonds on the Kagom\'e lattice correspond to
occupied bonds on the dice, and $p_c($dice$) = 1 - p_c($Kagom\'e).}

\figure{A tiling representation of the hull walk on the Kagom\'e lattice,
shown in the central part of the figure.
The tile marked with probability $p$ corresponds to an occupied bond on the underlying
Kagom\'e lattice, while the one marked $1-p$ corresponds to a vacant bond.}

\figure{The representation of the (6,4,3,4)-lattice on a rectilinear
grid, as utilized in the computer program's two-dimensional array.}

\figure{A plot of the estimate $p_c(g)$ determined by (6), versus the gradient $g$,
implying the linear relation (9).  The error bars represent one standard deviation
of statistical error.}

\figure{A similar plot as in Fig.~6,  with data from two systems (case A, 
dashed line through data points, and case B, solid line through data points)
in which the gradient is not locally uniform, as described in the Appendix.
The straight line represents the fit of the data of Fig.~6 for
the system the uniform gradient.  For small $g$, all systems follow the 
same asymptotic behavior.}

\end